\documentclass[sigconf]{acmart}

\setcopyright{none}               
\setcopyright{none}
\acmConference[]{To appear in Scored}{2026}{Preprint} 
\usepackage[utf8]{inputenc}
\usepackage{graphicx}
\usepackage{subcaption}
\usepackage[ruled,linesnumbered,commentsnumbered]{algorithm2e}
\usepackage{xcolor}
\usepackage{amsmath}
\usepackage{cleveref}
\usepackage{siunitx}
\usepackage{comment}
\usepackage{framed}
\usepackage{xspace}
\usepackage{float}

\definecolor{jblueleft}{RGB}{0, 79, 144}   
\definecolor{jblueinner}{RGB}{240, 248, 255} 

\newenvironment{conclusionBox}{%
    \MakeFramed{\advance\hsize-\width\FrameRestore}%
}{\endMakeFramed}

\newcommand{\conclusion}[1]{
\begin{conclusionBox}
    #1
\end{conclusionBox}
}

\newcommand{\crossfork}{cross-fork\xspace}

\newcommand{\commit}[1]{\texttt{\small #1}}

\newcommand{\upstreamrepo}{7162\xspace}

\newcommand{\impactedcommitshort}{158.9 million\xspace}

\newcommand{\impactedcommitinforkshort}{86.3 million\xspace}
\newcommand{\impactedcommitinforkpercentage}{54.3\%\xspace}

\newcommand{\unpatchedforkinheadshort}{1.7 million\xspace}

\newcommand{\unpatchedpairinhead}{53 496 067}
\newcommand{\unpatchedpairinheadshort}{53 million\xspace}

\newcommand{\impactedforkshort}{2.2 million\xspace}

\newcommand{\averageforkbyrepo}{432\xspace}
\newcommand{\medianforkbyrepo}{5\xspace}
\newcommand{\firstquartileforkbyrepo}{0\xspace}

\begin{document}

\newcommand{\forkTotal}{195\xspace}
\newcommand{\forkNonEmpty}{195\xspace}
\newcommand{\forkCategories}{41\xspace}

\newcommand{\forkMean}{4.76\xspace}
\newcommand{\forkMedian}{2\xspace}
\newcommand{\forkMin}{1\xspace}
\newcommand{\forkMax}{100\xspace}
\newcommand{\forkqone}{1\xspace}
\newcommand{\forkqthree}{3\xspace}

\newcommand{\forkhttpsgithubcomrhysdvimwasmCount}{100\xspace}
\newcommand{\forkhttpsgithubcomrhysdvimwasmPercent}{51.3\%\xspace}
\newcommand{\forkhttpsgithubcomrhysdvimwasmPercentInt}{51\%\xspace}
\newcommand{\forkhttpsgithubcomSnapchatKeyDBCount}{8\xspace}
\newcommand{\forkhttpsgithubcomSnapchatKeyDBPercent}{4.1\%\xspace}
\newcommand{\forkhttpsgithubcomSnapchatKeyDBPercentInt}{4\%\xspace}
\newcommand{\forkhttpsgithubcomMotorolaMobilityLLCkernelmsmCount}{5\xspace}
\newcommand{\forkhttpsgithubcomMotorolaMobilityLLCkernelmsmPercent}{2.6\%\xspace}
\newcommand{\forkhttpsgithubcomMotorolaMobilityLLCkernelmsmPercentInt}{3\%\xspace}
\newcommand{\forkhttpsgithubcomgoogleghostkernelCount}{5\xspace}
\newcommand{\forkhttpsgithubcomgoogleghostkernelPercent}{2.6\%\xspace}
\newcommand{\forkhttpsgithubcomgoogleghostkernelPercentInt}{3\%\xspace}
\newcommand{\forkhttpsgithubcomlinuxkernellabslinuxCount}{5\xspace}
\newcommand{\forkhttpsgithubcomlinuxkernellabslinuxPercent}{2.6\%\xspace}
\newcommand{\forkhttpsgithubcomlinuxkernellabslinuxPercentInt}{3\%\xspace}
\newcommand{\forkhttpsgithubcomfelipecgitCount}{4\xspace}
\newcommand{\forkhttpsgithubcomfelipecgitPercent}{2.1\%\xspace}
\newcommand{\forkhttpsgithubcomfelipecgitPercentInt}{2\%\xspace}
\newcommand{\forkhttpsgithubcomgitstergitCount}{4\xspace}
\newcommand{\forkhttpsgithubcomgitstergitPercent}{2.1\%\xspace}
\newcommand{\forkhttpsgithubcomgitstergitPercentInt}{2\%\xspace}
\newcommand{\forkhttpsgithubcomadafruitcircuitpythonCount}{4\xspace}
\newcommand{\forkhttpsgithubcomadafruitcircuitpythonPercent}{2.1\%\xspace}
\newcommand{\forkhttpsgithubcomadafruitcircuitpythonPercentInt}{2\%\xspace}
\newcommand{\forkhttpsgithubcomreMarkablelinuxCount}{4\xspace}
\newcommand{\forkhttpsgithubcomreMarkablelinuxPercent}{2.1\%\xspace}
\newcommand{\forkhttpsgithubcomreMarkablelinuxPercentInt}{2\%\xspace}
\newcommand{\forkhttpsgithubcomsonyxperiadevkernelCount}{4\xspace}
\newcommand{\forkhttpsgithubcomsonyxperiadevkernelPercent}{2.1\%\xspace}
\newcommand{\forkhttpsgithubcomsonyxperiadevkernelPercentInt}{2\%\xspace}
\newcommand{\forkhttpsgithubcomgitpodioopenvscodeserverCount}{3\xspace}
\newcommand{\forkhttpsgithubcomgitpodioopenvscodeserverPercent}{1.5\%\xspace}
\newcommand{\forkhttpsgithubcomgitpodioopenvscodeserverPercentInt}{2\%\xspace}
\newcommand{\forkhttpsgithubcomddnetddnetCount}{3\xspace}
\newcommand{\forkhttpsgithubcomddnetddnetPercent}{1.5\%\xspace}
\newcommand{\forkhttpsgithubcomddnetddnetPercentInt}{2\%\xspace}
\newcommand{\forkhttpsgithubcomhardkernellinuxCount}{3\xspace}
\newcommand{\forkhttpsgithubcomhardkernellinuxPercent}{1.5\%\xspace}
\newcommand{\forkhttpsgithubcomhardkernellinuxPercentInt}{2\%\xspace}
\newcommand{\forkhttpsgithubcomrockchiplinuxkernelCount}{3\xspace}
\newcommand{\forkhttpsgithubcomrockchiplinuxkernelPercent}{1.5\%\xspace}
\newcommand{\forkhttpsgithubcomrockchiplinuxkernelPercentInt}{2\%\xspace}
\newcommand{\forkhttpsgithubcompandarepandaCount}{3\xspace}
\newcommand{\forkhttpsgithubcompandarepandaPercent}{1.5\%\xspace}
\newcommand{\forkhttpsgithubcompandarepandaPercentInt}{2\%\xspace}
\newcommand{\forkhttpsgithubcomXQuartzxorgserverCount}{3\xspace}
\newcommand{\forkhttpsgithubcomXQuartzxorgserverPercent}{1.5\%\xspace}
\newcommand{\forkhttpsgithubcomXQuartzxorgserverPercentInt}{2\%\xspace}
\newcommand{\forkhttpsgithubcomgitforwindowsgitCount}{3\xspace}
\newcommand{\forkhttpsgithubcomgitforwindowsgitPercent}{1.5\%\xspace}
\newcommand{\forkhttpsgithubcomgitforwindowsgitPercentInt}{2\%\xspace}
\newcommand{\forkhttpsgithubcomespressifqemuCount}{2\xspace}
\newcommand{\forkhttpsgithubcomespressifqemuPercent}{1.0\%\xspace}
\newcommand{\forkhttpsgithubcomespressifqemuPercentInt}{1\%\xspace}
\newcommand{\forkhttpsgithubcombeckusqemustmCount}{2\xspace}
\newcommand{\forkhttpsgithubcombeckusqemustmPercent}{1.0\%\xspace}
\newcommand{\forkhttpsgithubcombeckusqemustmPercentInt}{1\%\xspace}
\newcommand{\forkhttpsgithubcomyiichinayiiCount}{2\xspace}
\newcommand{\forkhttpsgithubcomyiichinayiiPercent}{1.0\%\xspace}
\newcommand{\forkhttpsgithubcomyiichinayiiPercentInt}{1\%\xspace}
\newcommand{\forkhttpsgithubcomytdlpatchedytdlpatchedCount}{2\xspace}
\newcommand{\forkhttpsgithubcomytdlpatchedytdlpatchedPercent}{1.0\%\xspace}
\newcommand{\forkhttpsgithubcomytdlpatchedytdlpatchedPercentInt}{1\%\xspace}
\newcommand{\forkhttpsgithubcomalibabacloudkernelCount}{2\xspace}
\newcommand{\forkhttpsgithubcomalibabacloudkernelPercent}{1.0\%\xspace}
\newcommand{\forkhttpsgithubcomalibabacloudkernelPercentInt}{1\%\xspace}
\newcommand{\forkhttpsgithubcomgominigomipsCount}{2\xspace}
\newcommand{\forkhttpsgithubcomgominigomipsPercent}{1.0\%\xspace}
\newcommand{\forkhttpsgithubcomgominigomipsPercentInt}{1\%\xspace}
\newcommand{\forkhttpsgithubcomcloudflarezlibCount}{2\xspace}
\newcommand{\forkhttpsgithubcomcloudflarezlibPercent}{1.0\%\xspace}
\newcommand{\forkhttpsgithubcomcloudflarezlibPercentInt}{1\%\xspace}
\newcommand{\forkhttpsgithubcomapachenuttxCount}{1\xspace}
\newcommand{\forkhttpsgithubcomapachenuttxPercent}{0.5\%\xspace}
\newcommand{\forkhttpsgithubcomapachenuttxPercentInt}{1\%\xspace}
\newcommand{\forkhttpsgithubcomemercoinemercoinCount}{1\xspace}
\newcommand{\forkhttpsgithubcomemercoinemercoinPercent}{0.5\%\xspace}
\newcommand{\forkhttpsgithubcomemercoinemercoinPercentInt}{1\%\xspace}
\newcommand{\forkhttpsgithubcomgitgitCount}{1\xspace}
\newcommand{\forkhttpsgithubcomgitgitPercent}{0.5\%\xspace}
\newcommand{\forkhttpsgithubcomgitgitPercentInt}{1\%\xspace}
\newcommand{\forkhttpsgithubcomjenkinscihttprequestpluginCount}{1\xspace}
\newcommand{\forkhttpsgithubcomjenkinscihttprequestpluginPercent}{0.5\%\xspace}
\newcommand{\forkhttpsgithubcomjenkinscihttprequestpluginPercentInt}{1\%\xspace}
\newcommand{\forkhttpsgithubcomspreespreeauthdeviseCount}{1\xspace}
\newcommand{\forkhttpsgithubcomspreespreeauthdevisePercent}{0.5\%\xspace}
\newcommand{\forkhttpsgithubcomspreespreeauthdevisePercentInt}{1\%\xspace}
\newcommand{\forkhttpsgithubcomshyouheirubyCount}{1\xspace}
\newcommand{\forkhttpsgithubcomshyouheirubyPercent}{0.5\%\xspace}
\newcommand{\forkhttpsgithubcomshyouheirubyPercentInt}{1\%\xspace}
\newcommand{\forkhttpsgithubcomjlkomodoCount}{1\xspace}
\newcommand{\forkhttpsgithubcomjlkomodoPercent}{0.5\%\xspace}
\newcommand{\forkhttpsgithubcomjlkomodoPercentInt}{1\%\xspace}
\newcommand{\forkhttpsgithubcomgcashbchdCount}{1\xspace}
\newcommand{\forkhttpsgithubcomgcashbchdPercent}{0.5\%\xspace}
\newcommand{\forkhttpsgithubcomgcashbchdPercentInt}{1\%\xspace}
\newcommand{\forkhttpsgithubcomBitcoinABCbitcoinabcCount}{1\xspace}
\newcommand{\forkhttpsgithubcomBitcoinABCbitcoinabcPercent}{0.5\%\xspace}
\newcommand{\forkhttpsgithubcomBitcoinABCbitcoinabcPercentInt}{1\%\xspace}
\newcommand{\forkhttpsgithubcomKomodoPlatformkomodoCount}{1\xspace}
\newcommand{\forkhttpsgithubcomKomodoPlatformkomodoPercent}{0.5\%\xspace}
\newcommand{\forkhttpsgithubcomKomodoPlatformkomodoPercentInt}{1\%\xspace}
\newcommand{\forkhttpsgithubcomPirateNetworkpirateCount}{1\xspace}
\newcommand{\forkhttpsgithubcomPirateNetworkpiratePercent}{0.5\%\xspace}
\newcommand{\forkhttpsgithubcomPirateNetworkpiratePercentInt}{1\%\xspace}
\newcommand{\forkhttpsgithubcomsyscoinsyscoinCount}{1\xspace}
\newcommand{\forkhttpsgithubcomsyscoinsyscoinPercent}{0.5\%\xspace}
\newcommand{\forkhttpsgithubcomsyscoinsyscoinPercentInt}{1\%\xspace}
\newcommand{\forkhttpsgithubcomsyndbggoenvCount}{1\xspace}
\newcommand{\forkhttpsgithubcomsyndbggoenvPercent}{0.5\%\xspace}
\newcommand{\forkhttpsgithubcomsyndbggoenvPercentInt}{1\%\xspace}
\newcommand{\forkhttpsgithubcomjustmoonnodeextendCount}{1\xspace}
\newcommand{\forkhttpsgithubcomjustmoonnodeextendPercent}{0.5\%\xspace}
\newcommand{\forkhttpsgithubcomjustmoonnodeextendPercentInt}{1\%\xspace}
\newcommand{\forkhttpsgithubcombitcoinsvbitcoinsvCount}{1\xspace}
\newcommand{\forkhttpsgithubcombitcoinsvbitcoinsvPercent}{0.5\%\xspace}
\newcommand{\forkhttpsgithubcombitcoinsvbitcoinsvPercentInt}{1\%\xspace}
\newcommand{\forkhttpsgithubcomexporeactnativeCount}{1\xspace}
\newcommand{\forkhttpsgithubcomexporeactnativePercent}{0.5\%\xspace}
\newcommand{\forkhttpsgithubcomexporeactnativePercentInt}{1\%\xspace}
\newcommand{\forkhttpsgithubcombitnineossagensgraphCount}{1\xspace}
\newcommand{\forkhttpsgithubcombitnineossagensgraphPercent}{0.5\%\xspace}
\newcommand{\forkhttpsgithubcombitnineossagensgraphPercentInt}{1\%\xspace}
\newcommand{\AuthorEvalcountPOSITIVE}{135\xspace}
\newcommand{\AuthorEvalcountNEGATIVE}{60\xspace}
\newcommand{\AuthorEvalcountTOTAL}{195\xspace}
\newcommand{\AuthorEvalcountprecision}{0.69\xspace}
\newcommand{\AuthorEvalcountprecisionpercent}{69\xspace}
\newcommand{\AuthorEvalcountcorrectedprecision}{0.48\xspace}
\newcommand{\AuthorEvalcountmeanofprecisionbyfork}{0.59\xspace}
\newcommand{\AuthorEvalWithoutOutliercountPOSITIVE}{52\xspace}
\newcommand{\AuthorEvalWithoutOutliercountNEGATIVE}{35\xspace}
\newcommand{\AuthorEvalWithoutOutliercountTOTAL}{87\xspace}
\newcommand{\AuthorEvalWithoutOutliercountprecision}{0.6\xspace}
\newcommand{\AuthorEvalWithoutOutliercountprecisionpercent}{60\xspace}
\newcommand{\AuthorEvalWithoutOutliercountmeanofprecisionbyfork}{0.6\xspace}
\newcommand{\OwnerNotificationcountAFFECTED}{9\xspace}
\newcommand{\OwnerNotificationcountEXCLUDED}{8\xspace}
\newcommand{\OwnerNotificationcountNOTAFFECTED}{4\xspace}
\newcommand{\OwnerNotificationcountTOTALRESPONSE}{13\xspace}
\newcommand{\OwnerNotificationcountresponseprecision}{0.69\xspace}
\newcommand{\OwnerNotificationcountresponseprecisionpercent}{69\xspace}
\graphicspath{{../}}

\title[Detecting One-Day Vulnerabilities in Open-Source Forks with Global History Analysis]{Did You Forkget It? Detecting One-Day Vulnerabilities in Open-Source Forks with Global History Analysis}

 \author[Lefeuvre]{Romain Lefeuvre}
 \affiliation{
 \institution{University of Rennes}
 \city{Rennes}
 \country{France}}
 \email{romain.lefeuvre@inria.fr}
 \orcid{0009-0008-2019-1058}

 \author[Reux]{Charly Reux}
 \affiliation{
 \institution{University of Rennes}
 \country{France}
   \city{Rennes}}
 \email{charly.reux@inria.fr}
 \orcid{0009-0008-8708-8362}

 \author[Zacchiroli]{Stefano Zacchiroli}
 \orcid{0000-0002-4576-136X}
 \affiliation{
   \department{LTCI, Télécom Paris}
   \institution{Institut Polytechnique de Paris}
   \city{Palaiseau}
   \country{France}
 }
 \email{stefano.zacchiroli@telecom-paris.fr}
 
 \author[Barais]{Olivier Barais}
 \orcid{0000-0002-4551-8562}
 \affiliation{
 \institution{University of Rennes}
 \country{France}
   \city{Rennes}}
 \email{olivier.barais@irisa.fr}
 
 \author[Combemale]{Benoit Combemale}
 \orcid{0000-0002-7104-7848}
 \affiliation{
 \institution{Inria}
 \country{France}
   \city{Rennes}
}
\email{benoit.combemale@inria.fr}

\begin{abstract}

Tracking vulnerabilities inherited from third-party open-source software is a well-known challenge, often addressed by tracing the threads of dependency information. At scale, existing approaches precompute the vulnerable versions of software associated with known CVEs, based on declared impacted versions or using \emph{local history analysis}. 
However, vulnerabilities can also propagate through \emph{forking}: a repository forked after a vulnerability is introduced but before it is patched may remain vulnerable long after the original repository has been fixed. Existing \emph{history analysis} approaches analyze only the repository referenced by the CVE, providing a local view of the ecosystem that excludes forks sharing part of its development history. Vulnerabilities disclosed and patched elsewhere in a fork ecosystem may therefore persist as one-day (known but unpatched) vulnerabilities without fork maintainers’ awareness.
This paper proposes a \emph{global history analysis} approach that conforms to the evaluation semantics defined by the OpenSSF Open Source Vulnerability (OSV) format, while extending them from repository-local histories to the global commit graph of the open-source ecosystem. Leveraging the graph of public code captured by Software Heritage, our approach propagates vulnerability introduction and fix information across shared commit histories and performs automated impact analysis.
Starting from \upstreamrepo repositories containing vulnerable commits listed in the OSV.dev vulnerability database, we propagate vulnerability information to \impactedforkshort forks. We evaluate our approach on a sample of 195 $\langle\mathit{fork}, \mathit{vulnerability}\rangle$ pairs from popular repositories,
manually auditing their code and contacting their maintainers for confirmation and responsible disclosure. This process identified \AuthorEvalcountPOSITIVE high-severity one-day vulnerabilities, achieving a precision of \AuthorEvalcountprecision, with \OwnerNotificationcountAFFECTED cases confirmed by maintainers.

\end{abstract}

\maketitle
\section{Introduction} 
\label{sec:intro}

Open-Source Software (OSS) plays a critical role in the global digital infrastructure~\cite{hoffmann_value_2024}, its popularity having increased greatly over the past decades thanks to its unrivaled ease of code reuse.
OSS reuse is now a central practice in the software supply chain but can be considered a ``double-edged sword''~\cite{gkortzis2019double}, offering security benefits by involving a broader set of actors in maintaining code security, while also posing risks when vulnerabilities in reused software are not fixed in a timely manner or when reuse increases the attack surface (e.g., supply chain attacks). 

There are two main mechanisms for the reuse of OSS~\cite{Businge_22_maintenance_forks}: \emph{dependencies} and \emph{forks}.
The dependency mechanism involves calling into the code of a separate, freely licensed software via its API.
Forking means reusing the source code of an existing OSS to establish a separate development trajectory, resulting in distinct software~\cite{10.1145/3338906.3338918}. Forking serves multiple purposes: facilitating large-scale development through a GitFlow-like approach~\cite{rios2022gitworkflows}, addressing conflicts within the community, or extending functionality without integrating new ideas upstream~\cite{nyman2011fork, zhou_fork}.
This can be observed empirically: more than 40\% of repositories hosted on GitHub are forks\footnote{Analysis based on an export of Software Heritage as of April 3, 2024}.

The security of the global open-source ecosystem has attracted significant attention in recent years, particularly in the wake of high-profile software supply chain attacks on specific dependencies~\cite{10179304}. To address these growing risks, governments have introduced regulations requiring the software industry to identify and document the software supply chain. In the United States, Executive Order 14028~\cite{EO14028_2021} pushes for the use of Software Bills of Materials (SBOMs), defined as ``formal records containing the details and supply chain relationships of the various components used in building software.'' Similarly, the European Union's Cyber Resilience Act (CRA)~\cite{cra2024} strengthens the security standards for digital products and holds ``manufacturers'' of digital services responsible for security issues in their supply chain, including those arising from reused software. These regulations place responsibility on service manufacturers to conduct appropriate audits and remediate vulnerabilities related to reused code.

Most research in this area has focused on the challenges associated with secure dependency management~\cite{he2023automating}.
The risk of vulnerability propagation through dependencies is well documented in the literature, which emphasizes the need for thorough audits to identify vulnerabilities and implement mitigation measures, such as upgrading to a secure version (or applying the corresponding patches) as quickly as possible. The need to assess dependency-related vulnerabilities has driven the development of tools such as GitHub Dependabot~\cite{GitHubDependabot} and OSV-Scanner~\cite{GoogleOSVScanner}, which leverage vulnerability databases built from either manual annotations or \emph{history analysis}. 

The security implications of fork-based reuse have received comparatively less attention, in stark contrast to how widespread and foundational fork-based reuse is in major OSS: the Linux kernel, for instance, has numerous forks, including prominent derived kernels such as Android.

\paragraph{Problem statement}
This paper focuses on the underexplored topic of ``\crossfork'' vulnerability propagation in the open-source ecosystem.
Forks share parts of a common code history, and a vulnerability identified in one fork is likely to affect others, depending on when forking happened and how changes in the initial (``upstream'') code repository are incorporated into (``downstream'') forks. 
For example, if a downstream code repository is forked after an upstream commit that introduces a ``high severity'' vulnerability and before the corresponding commit that fixes it (assuming it already exists), the downstream code repository will most likely remain vulnerable until a fixing commit is merged back there.
From the moment the vulnerability is discovered and the fix commit is integrated, the downstream code repository is affected by a so-called \emph{one-day vulnerability}: a known, but unpatched vulnerability.

Existing history analysis approaches, used to identify the vulnerable software versions impacted by a CVE, analyze only the repository referenced by the CVE, providing a local view of the ecosystem that excludes forks sharing part of its development history. Vulnerabilities disclosed and patched elsewhere in a fork ecosystem may therefore persist as one-day (known but unpatched) vulnerabilities.

Given the current lack of approaches and tools to track cross-fork vulnerability propagation, fork maintainers are tasked with manually tracking vulnerabilities associated with their fork ecosystem and assessing, for each vulnerability, whether it applies to them and if the corresponding fixes have been integrated.
Fork users are also impacted, due to the fact that only major, well-known forks are tracked by vulnerability databases: users of the myriad other forks will generally be unaware that they are impacted by upstream one-day vulnerabilities inherited via forking.

Addressing this issue requires: (1) identifying forks, ideally at a global scale; (2) keeping track of which commits introduce and fix security vulnerabilities; and (3) propagating the information about which commits are vulnerable, from upstream OSS to downstream forks.
This challenge is compounded by the diverse and heterogeneous nature of software forges (GitHub, GitLab in multiple self-hosted instances\ldots) and the use of different version control systems (Git, Mercurial, Subversion\ldots). To tackle this challenge, we propose shifting from repository-local history analysis to a global history analysis of the open-source ecosystem. Rather than analyzing each repository in isolation, our approach operates on a representation of the global commit graph that connects repositories through their shared development histories, enabling vulnerability information to be propagated across entire fork ecosystems.

\paragraph{Contributions}

This paper introduces the following contributions: 
\begin{itemize}

    \item A \emph{global history analysis} approach that conforms to the evaluation semantics defined by the OSV format, while extending them from repository-local histories to the global commit graph of the open-source ecosystem, thereby enabling fork maintainers to identify one-day vulnerabilities by tracking introducing and fixing commits across fork ecosystems.
    \item An implementation of the approach that can scale to the complete Software Heritage commit graph (consisting of more than 5 billion unique commits), enabling commit-level vulnerability tracking across heterogeneous public forges and version control systems.
    \item A large-scale study showing that, starting from \upstreamrepo repositories, referenced in the OSV database as having been affected by vulnerabilities in the past, vulnerabilities propagate to \impactedforkshort potentially impacted forks.
    \item The evaluation of the approach on a sample of real, high-impact one-day vulnerabilities in independent forks. After filtering for significant use (based on GitHub stars) and severity (based on CVSS scores), we identified \AuthorEvalcountPOSITIVE{} cases of vulnerabilities (precision \AuthorEvalcountprecision{}) and obtained further positive confirmation from maintainers for \OwnerNotificationcountAFFECTED{} high-severity one-day vulnerabilities.
\end{itemize}

\paragraph{Data availability statement}

A replication package of the experiments presented in this paper is available online~\cite{DidYouForkgetItReproductionAnon}.

\section{Background}
\label{sec:background}

\paragraph{Forking}
\label{sec:forking}
This open-source practice consists in creating a new code repository from an existing one, usually preserving prior development history. Traditionally, ``forking'' meant splitting development efforts to bring an OSS in a different technical direction than intended by current maintainers, resulting in distinct software, referred to as a \emph{hard fork}.
With the advent of collaborative code hosting platforms~\cite{dabbish2012socialcoding}, a new kind of fork has emerged, called \emph{social fork}: contributors create personal forks of an OSS to 
propose changes via pull/merge requests~\cite{zhou_fork}. Social forks tend to be ephemeral, but some grow their own user base, 
blurring the lines between hard and social forks.

In this paper, to \emph{discover forks} of an OSS, we rely on the structural properties of the global commit graph produced by distributed version control systems (DVCSs) like Git, using the definition of \emph{shared commit fork} from Pietri et al.~\cite{fork}: {``\itshape a code repository $B$ is a shared commit fork of code repository $A$, [\ldots] if there exists a commit $c$ contained in the development histories of both $A$ and $B$.''} This notion is undirected---$A$ is a fork of $B$ \emph{and vice-versa} as soon as they share a commit---so we refer to a \emph{fork ecosystem} as a group of repositories that are all forks of each other. This definition is broader than the forge-specific ``fork'' relation, and enables detecting forks that were not created through the ``fork button'', including repositories pushed independently and ``cross-forge'' forks hosted on different platforms (e.g., a GitLab.com fork of a GitHub repository).

\paragraph{Detection of vulnerabilities induced by code reuse}
Analyzing dependencies at scale is now common practice with tools such as GitHub Dependabot~\cite{GitHubDependabot} and OSV-Scanner~\cite{GoogleOSVScanner}: they first build a database mapping known vulnerabilities to affected software versions, then analyze a given repository's dependencies against that database.
Ideally, static or dynamic detection techniques (cf. \Cref{sec:related}) would identify affected versions for the database, but these are often language-specific and too resource-intensive to apply at scale, as they typically require file-level analysis.
Instead, many tools rely either on manually declared affected-version lists in CVE records, or on techniques that mine version control history to automatically identify vulnerable versions (e.g., OSV). We refer to the latter as \emph{history analysis techniques}: unlike static or dynamic analysis, they rely solely on the commit graph, identifying versions that contain the vulnerability-introducing commit without the corresponding patch commit, which lets them scale efficiently by avoiding file-level analysis.

\paragraph{OSV (Open-Source Vulnerabilities)}
\label{sec:osv}
This OpenSSF project\footnote{\url{https://osv.dev}} consists of (1) a standard format\footnote{\url{https://ossf.github.io/osv-schema}} to describe vulnerabilities affecting OSS, and (2) a public database of vulnerability advisories in that format, aggregating security information from more than 20 software ecosystems. The OSV format associates vulnerabilities with \emph{version ranges}, described as sequences of \emph{events}. For example, an \emph{introduced} event refers to the version where the vulnerability was introduced, and a \emph{fixed} event refers to the version containing the patch. The format also defines events related to practical requirements, such as \emph{last\_affected}, which is used to reference the last known version with the vulnerability when the exact patch version is not known. 
Versions can be specified as version numbers or Git commit SHA1s; reasoning at the commit level enables more fine-grained analyses than version numbers and is particularly useful when studying forks. In the remainder of the paper, we only consider OSV vulnerabilities whose ranges are described using Git ranges.

The OSV database backs an API that answers queries like \emph{``Is version X of software Y affected by any known vulnerability?''}, where ``version'' can likewise be a Git commit identifier. To answer such queries, OSV periodically crawls the repositories associated with a vulnerability and evaluates an algorithm (part of the OSV standard), maintaining an up-to-date $\mathit{commit}\to^{*}\mathit{vulnerability}$ relationship. Crucially, OSV only knows about commits in the upstream repository, and would produce false negatives when queried about commits only found in downstream forks.

\paragraph{Software Heritage (SWH)}
\label{sec:swh}
The largest public archive of software source code and associated development history~\cite{di2017software}, SWH archives code from more than 400 million OSS projects coming from development forges (GitHub, GitLab, etc.) and package manager repositories, comprising to date more than 20 billion files and 5 billion commits\footnote{\url{https://www.softwareheritage.org}}.
The SWH data model is a deduplicated Merkle directed acyclic graph (DAG) structure~\cite{merkle_dag}, with nodes representing source code artifacts, including (Git) commits and releases, each identified by a unique and persistent ``SoftWare Hash IDentifier'' (SWHID)\footnote{\url{https://swhid.org}, ISO/IEC standard 18670:2025} that is, for commits, mutually derivable with the corresponding Git identifier.
The SWH graph, the best current approximation of the global commit graph, is formed by around 50 billion nodes and 0.5 trillion arcs, and can be loaded into main memory using a compressed representation~\cite{swh_graph} that allows scale-up analyses that would be prohibitively costly in other representations.

In the SWH graph, repositories are represented as \emph{origins} associated with \emph{snapshots}, which reference commits corresponding to branch heads.
The \emph{deduplication} property of the SWH graph makes it particularly useful for studying forks: forks hosted on different platforms, as well as those independently pushed to the same platform, become part of the same connected component of the graph, allowing forks to be studied globally, as previously done in~\cite{fork}.

\section{Tracking vulnerable commits across fork ecosystems}
\label{sec:approach}

We propose to track commits that introduce and fix known vulnerabilities at a global-scale, rather than on a per-repository basis.
We postulate, and later verify experimentally, that this novel approach will let us identify forks of open-source projects affected by one-day vulnerabilities, unknown to their maintainers and other downstream stakeholders in the open-source software supply chain.

\subsection{Challenge}
\label{sec:challenge}

Current \emph{history analysis} approaches for identifying specific commits affected by a vulnerability rely on local (repository-level) analysis.
For example, in the case of \url{OSV.dev}, the Git history of each code repository associated in the past with at least one vulnerability is cloned and analyzed separately to detect vulnerable commits.
A database of vulnerable commits is then used as a backend for an API that allows looking up vulnerabilities associated with a given commit.
Forks of OSS share part of a common development history with their parent code repository, potentially inheriting unfixed vulnerabilities.
Local \emph{history analyses} (i.e., analyses performed on a local model, such as an individual Git repository, as opposed to a global model, such as the Software Heritage (SWH) archive), induce a ``blind spot'' problem that can result in one-day vulnerabilities being overlooked in forks.

Let us look at a real case of an unpatched fork involving a vulnerability with a high severity on the Common Vulnerability Scoring System (CVSS), identified as CVE-2019-13164,\footnote{\url{https://nvd.nist.gov/vuln/detail/CVE-2019-13164}} impacting PANDA, a fork of QEMU.
QEMU~\cite{bellard2005qemu} is a well-known open-source generic machine emulator and virtualizer, whereas ``PANDA is an open-source platform for architecture-Neutral dynamic analysis [\ldots] built upon the QEMU whole system emulator.''\footnote{\url{https://panda.re}}
PANDA started as a hard fork of QEMU and then evolved independently for over a decade, rarely integrating commits from the upstream code repository.
Owing to their initially shared development history, PANDA inherits vulnerabilities present in QEMU at the time of the fork, including CVE-2019-13164. 
\begin{figure}
\centering
\includegraphics[width=\linewidth]{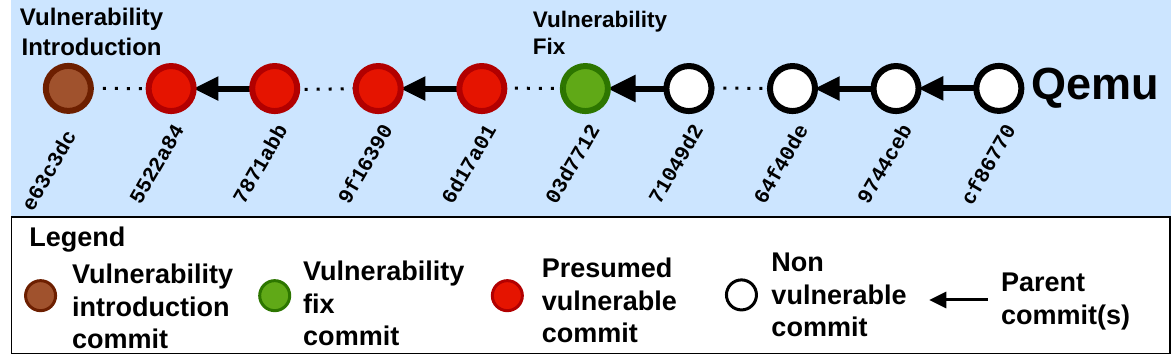}
\caption{QEMU Git history: local repository-level analysis. \small{Note that commits are built bottom-up on previous ones. Most recent commits are hence graph roots, shown on the right in the picture.}}
\Description{QEMU Git history: local repository-level analysis. Note that commits are built bottom-up on previous ones. Most recent commits are hence graph roots, shown on the right in the picture.}
\label{fig:A}
\end{figure}
\begin{figure}
    
\centering
\includegraphics[width=\linewidth]{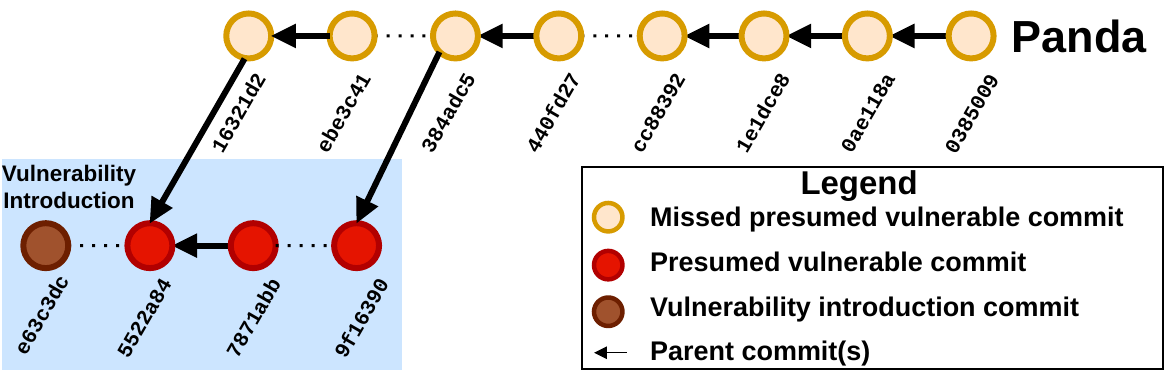}
 \captionsetup{}
\caption{Propagation of vulnerable commits from upstream QEMU repository to its downstream PANDA fork.}
\Description{Propagation of vulnerable commits from upstream QEMU repository to its downstream PANDA fork.}
\label{fig:B}
\end{figure}

\Cref{fig:A} illustrates the vulnerable commits in QEMU Git history, identified with a local analysis such as \url{OSV.dev}.
In this situation, the QEMU Git repository, referenced in the OSV vulnerability report,  is cloned and commits between the introduction and fix commit are listed and considered as vulnerable.
This specific vulnerability is considered to be already present in the initial Git commit of QEMU (\commit{e63c3dc}) and was eventually fixed in commit \commit{03d7712}.

\Cref{fig:B} depicts a portion of the PANDA Git history, which does not include the fix commit \commit{03d7712}.
The portion of the history highlighted in the blue box is shared with QEMU. The commits in this portion are identical to those in the QEMU repository, as indicated by their matching Git identifiers. Therefore, these commits are already considered vulnerable with QEMU local analysis (cf. \Cref{fig:A}). In contrast, the subsequent portion of PANDA's Git history contains vulnerable commits that are exclusive to PANDA and, therefore, lie entirely beyond the scope of QEMU's local analysis.
Neither is the PANDA repository analyzed separately: the OSS is not referenced in the associated vulnerability report, leading to an overlooked one-day vulnerability.
Note that the \emph{new} vulnerable commits in PANDA history have commit identities (as SHA1 identifiers) missing from the upstream QEMU repository.
Therefore they cannot be spotted as vulnerable based on commit identities only, because they will not be found in vulnerability databases which only crawl the original repositories or, at best, noteworthy hard forks.

Handling these cases requires a two-step process: first, identify all forks that may be impacted by a vulnerability, and second, analyze each of them.
However, identifying all forks of a code repository requires systematically examining open-source repositories across multiple code-hosting platforms.
Even if a comprehensive list of public forks of a given vulnerable upstream code repository could be obtained, analyzing each fork individually would lead to significant redundancy in the effort.
Forking a code repository is cheap, whereas vulnerability tracking is resource-intensive.

\subsection{Approach}

\begin{figure}
\centering
\includegraphics[width=1\linewidth]{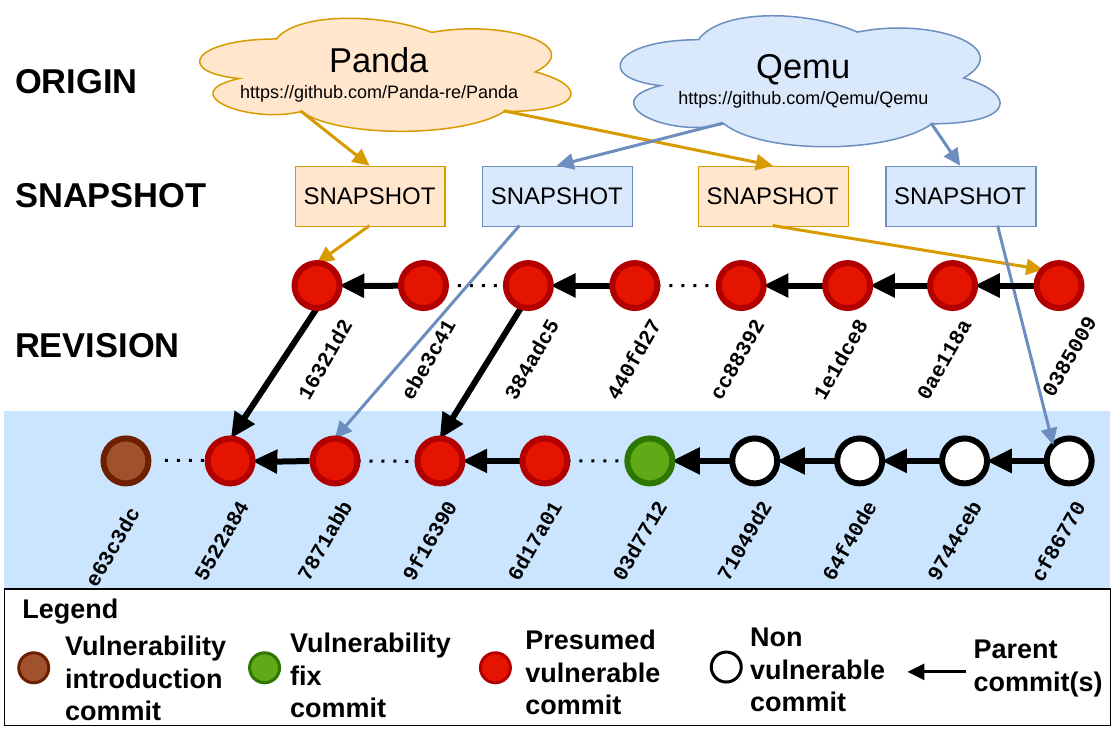}
\caption{Running example on SWH model}
\Description{Running example on SWH model}

\label{fig:global}

\end{figure}

To address this challenge, we introduce a new \textit{history analysis} approach based on a global model of commits, such as the one materialized by SWH (see \Cref{sec:swh}).
In this model, code histories are deduplicated and thus linked together through the identities of shared commits.   \Cref{fig:global} recasts our QEMU/PANDA example in the context of a global commit model (here, SWH). The list of vulnerable commits can now be identified by listing commit nodes between introduction and fix commits.
As a result, all commits in PANDA that were not detectable as vulnerable by separately analyzing the two repositories can now be spotted: by propagating the information that commit \commit{e63c3dc} introduced the vulnerability and that commit \commit{03d7712} fixed it, we can detect all PANDA commits starting at commit \commit{16321d2} as potentially vulnerable.

\label{sec:algo_sec}

To implement this idea in practice, we need to ``label'' each commit in the global commit graph with the set of known vulnerabilities affecting it.
We obtain these labels by propagating commit-level information about which commits introduce or fix vulnerabilities from upstream repositories to downstream forks.
Our algorithm formalizes the evaluation semantics of OSV,\footnote{\url{https://ossf.github.io/osv-schema/\#evaluation}} which is applied in that context to individual repositories, and generalizes it to the global commit graph.

Given a vulnerability range $R$, we perform a depth-first traversal of the global commit graph starting from the introducing (\texttt{intro}) commits, propagating a \texttt{patched}/\texttt{unpatched} status forward to every reachable commit: a commit is marked patched as soon as it is itself a fixing (\texttt{fixed}) commit, or when any of its parents is already patched (so a merge commit is patched if at least one of its parents is).
Beyond the \texttt{intro} and \texttt{fixed} events used here, the OSV standard also defines edge-case events---\texttt{last\_affected} and \texttt{limit}---whose handling, together with reintroductions of the same vulnerability along different branches, follows the OSV evaluation semantics that our algorithm generalizes to the global commit graph.
This process is repeated independently for each vulnerability range, and the results are aggregated into a global mapping from commits to the vulnerabilities affecting them.
Although identifying vulnerable commits between introduction and fix may seem straightforward, this formalization is needed to correctly handle the full range of event types defined by the OSV standard. 
An implementation of our approach is available in the replication package~\cite{DidYouForkgetItReproductionAnon}.

We described our automated approach for propagating commit-level vulnerability information from a code repository history to its forks using a unified model at the global scale of all publicly available commits. Before using this approach to look for actual one-day vulnerabilities (\Cref{sec:oneday}), we first need to check that it is practically feasible to propagate introduction/fix information across the global commit graph despite its size---from around 100K--1M commits for a single repository to billions of commits at the global scale.
Once this is established, the propagation of introduction/fix information leads to the identification of \emph{potentially} vulnerable commits in all forks of a given open-source project. Throughout this paper, \emph{vulnerable} denotes the absence of a known patch, not demonstrated exploitability or code-path reachability, which we do not test. This label remains only \emph{potential} at this stage, as forks that contain introducing, but not fixing, commits may still be unaffected in practice, for reasons we explore in \Cref{sec:oneday}.

\subsection{Experimental protocol}
\label{sec:label_protocol}
\begin{figure}[h]
    \centering
    \includegraphics[width=0.96\linewidth]{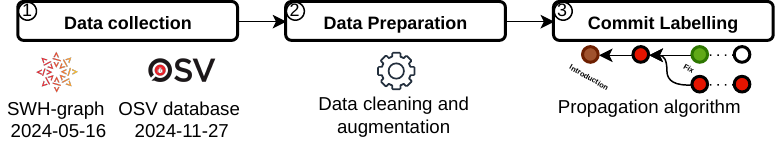}
    \caption{Overview of the experimental protocol steps for global propagation}
    \Description{Overview of the experimental protocol steps for global propagation}
    \label{fig:protocol_step_rq1}

\end{figure}

To validate this propagation at scale, we execute a three-step experimental protocol to propagate OSV vulnerability information across the global commit graph from Software Heritage.
\Cref{fig:protocol_step_rq1} presents an overview of the three steps of our experimental protocol.

\paragraph{Step 1: Data Collection.}
We retrieve two key datasets: (1) a recent export of the global commit graph from Software Heritage\footnote{\url{https://docs.softwareheritage.org/devel/swh-dataset/graph/dataset.html}.} (version 2024-05-16), and (2) vulnerability information exported from OSV\footnote{\url{https://google.github.io/osv.dev/data/\#data-dumps}} as of 2024-11-27.
\paragraph{Step 2: Data Preparation.}
We perform data cleaning and augmentation on the OSV dataset.
We identify and exclude ranges with inconsistencies, such as multiple event types pointing to the same commit or formatting issues.
We also exclude ranges containing events that reference commits absent from the SWH graph (e.g., commits more recent than our dataset snapshot).
For data augmentation, we handle two special cases.
First, when a vulnerability range specifies ``0'' as the introduction event (indicating the vulnerability existed from the repository's inception), we identify the leaf commits of the associated repositories to complete the range.
Second, we detect cherry-picked events across branches (potentially spanning different forks) by identifying commit messages containing the default Git pattern \texttt{``cherry picked from commit\ldots''}.
We add these detected cherry-picked commits as additional events within the same range.

\paragraph{Step 3: Commit Labeling.}
We apply the propagation algorithm described in \Cref{sec:algo_sec} to the SWH commit graph using the augmented OSV ranges.
The output is a comprehensive mapping from all SWH commits to their associated OSV vulnerability ranges.
All experiments were run on a server with  $16 \times 32$\,GiB DDR4 memory, 3200\,MHz; 6\,TiB of SSD storage; $2 \times$ AMD EPYC 7543 32-core CPUs, 2.8\,GHz;  Ubuntu 22.04.

\subsection{Results}
\label{sec:label_result}

\begin{figure}[htbp!]
    \centering
     \captionsetup{}

    \begin{minipage}[t]{0.46\textwidth}
        \vspace{0pt} 
        \centering
        \begin{subfigure}[t]{\textwidth}
            \includegraphics[width=\textwidth]{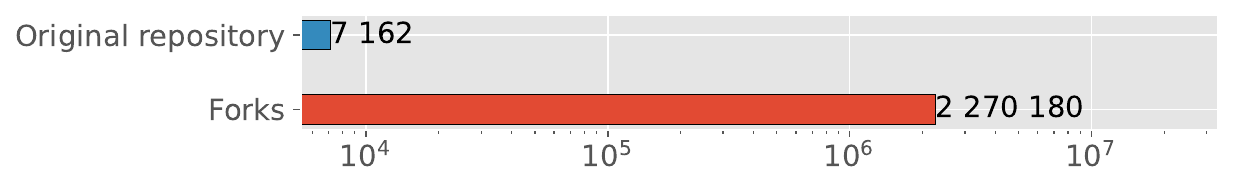}
            \caption{Number of forks compared to upstream repositories}
            \label{fig:fork_nb}
        \end{subfigure}

        \vspace{2mm} 

        \begin{subfigure}[t]{\textwidth}
            \includegraphics[width=\textwidth]{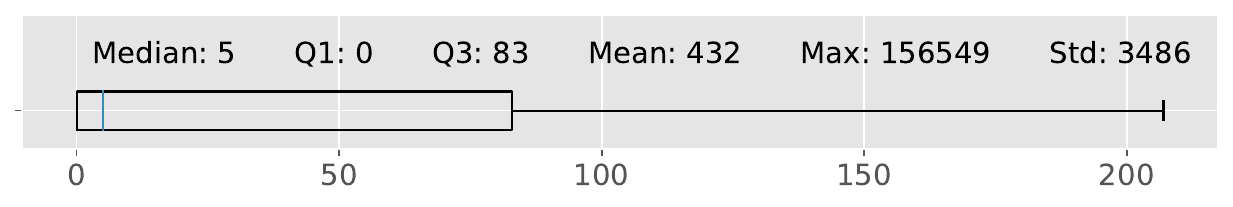}
            \caption{Distribution of the number of forks, by upstream code repository}
            \label{fig:fork_nb_distribution}
        \end{subfigure}

        \begin{subfigure}[t]{\textwidth}
            \includegraphics[width=\textwidth]{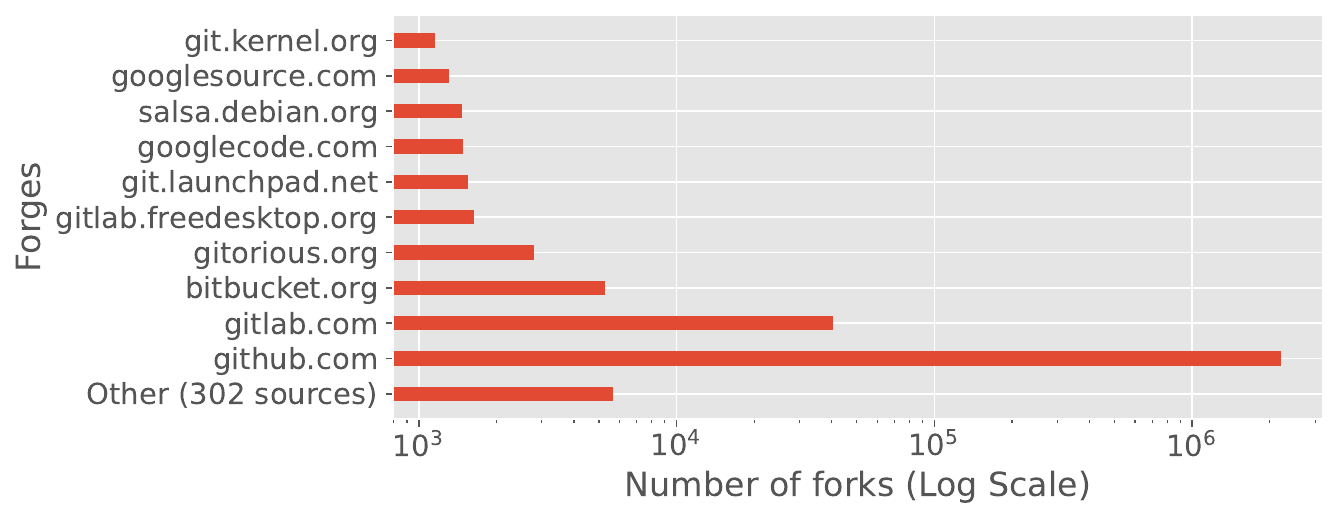}
            \caption{Number of forks by forge}
            \Description{Number of forks by forge}
            \label{fig:top_forge}
        \end{subfigure}
    \end{minipage}
    \hfill

    \caption{Forks of OSV-referenced repositories, having at least one new vulnerable commit.}
    \label{fig:fork}
\end{figure}

\Cref{fig:fork} shows that, after commit labeling (step 3), starting from a set of \upstreamrepo repositories referenced by OSV, we identified \impactedforkshort forks containing at least one presumed vulnerable commit.
Note that we filter forks to keep only those containing at least one presumed vulnerable commit that is missing from the upstream repository.
This excludes the trivial cases where downstream forks have not diverged from upstream in ways that are relevant for one-day vulnerabilities.
Such cases can already be handled with state-of-the-art history analysis approaches, and are hence uninteresting for our experimental validation.

On average, each upstream repository affected at some point in its development history by at least one known vulnerability propagates it to \averageforkbyrepo forks, which might remain affected.
A first quartile of \firstquartileforkbyrepo and a median of \medianforkbyrepo suggest that a few upstream repositories have a large number of potentially vulnerable forks.
Although the majority of vulnerable forks are hosted on GitHub, potentially vulnerable commits are spread across 312 distinct forges (by Internet domain name), confirming that vulnerable forks tend to spread around: platform-specific solutions are not enough to track them all.

\begin{figure}[h!]
 \captionsetup{}

\centering
 \includegraphics[width=\linewidth]{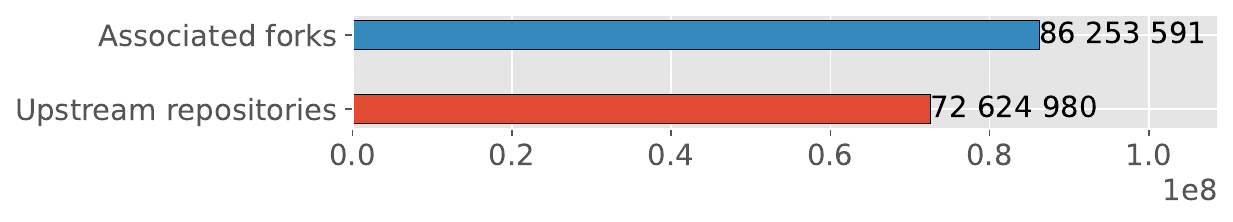}
        \caption{Number of new vulnerable commits in forks compared to upstream repositories}
        \Description{Number of new vulnerable commits in forks compared to upstream repositories}
        \label{fig:commits}
\end{figure}

As shown in \Cref{fig:commits}, we identified \impactedcommitshort commits as potentially vulnerable, \impactedcommitinforkshort of which can only be found in fork histories (by commit identity), accounting for \impactedcommitinforkpercentage of all potentially vulnerable commits.

\vspace{1em}
\conclusion{
\textbf{Takeaway:} Starting from \upstreamrepo repositories, our \emph{global history analysis} approach scales to the global commit graph, enabling tracking of vulnerability introductions and fixes across \impactedforkshort forks on 312 forges.

}

\section{Detecting one-day vulnerable forks}
\label{sec:oneday}

In the previous section, we propagated vulnerability information across the global commit graph. In this section, we evaluate our \emph{global history analysis} in the scenario of detecting \emph{one-day vulnerable forks}: forks that remain vulnerable in their latest revision to a vulnerability already patched elsewhere in their fork ecosystem.

Our \emph{global history analysis} propagated vulnerability information across \impactedforkshort forks, identifying \impactedcommitinforkshort \emph{potentially} vulnerable commits not present in upstream repositories.
However, the presence of a vulnerable commit in the development history of a fork does not necessarily mean the fork is vulnerable in practice, for three reasons.
First, the presumed vulnerable commit might not be the most recent (or ``head'') commit, to which users are most exposed.
Second, forks may have applied a patch \emph{equivalent} to the upstream fix through alternative means: they do fix the vulnerability, but are not recognized as doing so by our global history analysis.
Third, forks might not actually be vulnerable if they have diverged significantly from the upstream codebase, making it impossible to trigger the original vulnerable code path.

To account for these limitations, we evaluate our approach on a sample of actively used forks affected by high-impact vulnerabilities. This sampling strategy serves two purposes: it maximizes the likelihood of receiving a response from maintainers, on which we rely to build our ground truth; and it evaluates our tool in the context it is intended for, namely helping maintainers of hard forks with real user bases who could be impacted by an undisclosed one-day vulnerability.

\subsection{Research question}

We proceed to investigate whether the \emph{potential} vulnerabilities identified by our global propagation correspond to actual one-day vulnerabilities in real-world forked OSS. Thus, we introduce our research question:
\textbf{RQ: Can we leverage the proposed \emph{global history analysis} approach to detect at scale one-day vulnerabilities in real-world forked OSS?}

The next section presents the experimental protocol we used to answer this research question.

\subsection{Experimental protocol}
\label{sec:rq2_protocol}
To answer this RQ, we propose a multi-stage experimental protocol composed of three phases: first, we identify forks with unpatched head commits that remain vulnerable; second, we apply a series of filters to focus on high-impact forks and vulnerabilities; third, we validate our findings through both manual code audit and confirmation from project maintainers.
\begin{figure}
    \centering
    \includegraphics[width=0.96\linewidth]{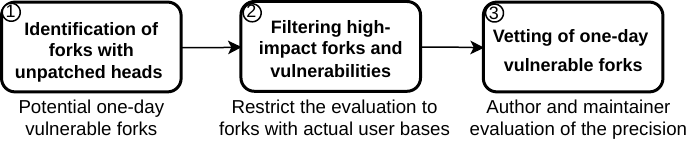}
    \caption{Overview of the experimental protocol steps}
    \Description{Overview of the experimental protocol steps}
    \label{fig:protocol_step_rq2}

\end{figure}
\Cref{fig:protocol_step_rq2} presents an overview of the three steps of our experimental protocol. 
\subsubsection{Identifying forks with unpatched heads}
To identify one-day vulnerabilities, we focus on the most recent commit in a repository branch that remains vulnerable, which we call \emph{unpatched head}. A fork containing an unpatched head is a \emph{presumed one-day vulnerable fork}. This allows us to exclude forks that had vulnerable commits at some point in their history, but have since patched them.
We further restrict this analysis to vulnerabilities specifying a \texttt{fixed} event: unlike \texttt{last\_affected}, which only bounds the affected range without referencing the patch itself, a \texttt{fixed} event gives us a commit reference, which is necessary to evaluate whether the fix has since been incorporated into a given fork.

\subsubsection{Filtering high-impact forks and vulnerabilities}
We identified in the previous section a large number of forks impacted in their history by referenced vulnerabilities (cf.~\Cref{fig:fork_nb}).
However, many of these are likely inactive or social forks with no real user base~\cite{zhou_fork, Kalliamvakou_peril_2016}.
To answer this RQ, evaluating our approach for detecting real-world one-day vulnerabilities requires us to identify \emph{high-impact forks}, which we define as forks with actual user bases and affected by high-severity vulnerabilities.

\paragraph{Popularity filtering}
We first filter on the size of the user-base of a forked repository.
To approximate this criterion, we rely on repository popularity metrics, specifically the number of stars and forks (of the fork itself, i.e., transitive forks of the upstream project) provided by GitHub.
The filtering criteria were defined as follows: we only keep forks hosted on GitHub with more than 100 stars and more than 10 forks. Note that, while Software Heritage (SWH) is more general than just GitHub, from now on we restrict the analysis to only projects archived by SWH and originally hosted on GitHub. This is because SWH currently does not archive popularity metrics for platforms other than GitHub.
Since the vast majority (97\%, as shown in \Cref{fig:top_forge}) of the forks detected as potentially vulnerable by our approach are hosted on GitHub anyway, this restriction has a limited quantitative impact. 

\paragraph{Severity filtering}
Second, we limit our analysis to vulnerabilities with CVSS severity of ``high'' or more (i.e., a CVSS base score $\geq 7.0$).
This allows us to focus on a limited number of potentially high-impact vulnerabilities, which we can in the following manually audit and contact maintainers about.
(It would be unfeasible to do so for vulnerabilities of all severities identified thus far.)

\paragraph{Stale filtering}
We exclude from further analysis forks that have been archived on GitHub or where the affected head commit is not located in the main branch.
The former denote products that are explicitly marked as no longer maintained; the latter are in general stale branches of limited impact on actual users.
We also exclude repositories with no recent activity, i.e., a most recent commit with date older than 2023-01-01. Finally, we exclude cross-referenced pairs  of $\langle\mathit{fork}, \mathit{vulnerability}\rangle$, i.e., forks that did not apply a specific vulnerability patch but have another vulnerability entry associated with a different patch.

\paragraph{Divergence filtering}
Forks can diverge in ways that make a specific vulnerability not applicable to them.
We use a heuristic to identify and exclude automatically divergent forks.
We examine the files impacted by the fix commit.
If at least one of the modified files is not present in the fork repository, we exclude the $\langle\mathit{fork}, \mathit{vulnerability}\rangle$ pair from further analysis.
The rationale for this is that if the code that needs patching is no longer there, the code base is unlikely to be affected by the associated vulnerability.
While executing this filtering step, we track file renames properly using \texttt{git log --follow --find-renames}.

\subsubsection{Vetting of one-day vulnerable forks}
\label{sec:vetting}
Once we have filtered the forks and vulnerabilities of interest, we evaluate the extent to which our approach successfully identifies one-day vulnerable forks.
We employ a two-phase validation procedure: In the first phase, we manually audit the source code for each $\langle\mathit{fork}, \mathit{vulnerability}\rangle$ pair, to identify potential false positives.
In the second phase, we contact the corresponding fork maintainers to obtain further confirmation from them regarding the impact of the vulnerability. This also acts as responsibly disclosing the vulnerability to them, so that they can fix it if needed.
In contrast to the general vulnerability detection literature, the detection of vulnerability propagation across fork ecosystems lacks a ground-truth dataset. We therefore treat maintainers’ responses as the ground truth for evaluating the proportion of false positives produced by our approach, i.e., the maintainer-confirmation rate (the fraction of pairs flagged as vulnerable that are genuinely vulnerable, $\mathit{TP}/(\mathit{TP}+\mathit{FP})$). However, in the absence of a comprehensive ground-truth dataset, we cannot reasonably estimate the proportion of false negatives (i.e., recall). Constructing a reliable ground-truth dataset for evaluating vulnerability propagation across fork ecosystems remains an open research problem.

\paragraph{Manual code audit}

Before manually evaluating each $\langle\mathit{fork},\\\mathit{vulnerability}\rangle$ pair, we conduct an evaluation of the  data quality of the vulnerability report used as input of our approach.
CVE databases are growing massively and rely heavily on unstructured data for which quality and consistency are a common concern~\cite{dong_vuln_report}.
Although it follows a standardized format, the OSV database—which aggregates multiple vulnerability databases—is not exempt from data-quality issues.
For instance, many ``fixed'' commits do not precisely reference the actual vulnerability patches.
Instead, fixed commits may point to catch-all release commits that include all changes associated with a new release, making it difficult to pinpoint the actual change that carried the security fix (cf.~\Cref{sec:threats}).
Since these imprecise entries make it impossible to reliably assess whether a fork has incorporated the corresponding vulnerability patch, and thus impede the evaluation of our approach, we remove them from our analysis.
In addition, the objective of this evaluation is to assess our approach, and not the quality of CVE reports. 

Then, since the objective is to evaluate our approach rather than its current prototype implementation, we complement the global cherry-pick detection heuristic with a finer-grained one based on Git patch IDs.\footnote{\url{https://git-scm.com/docs/git-patch-id}}
While we already detected and propagated some cherry picks during the global propagation described in \Cref{sec:label_protocol}, we did so solely based on commit message traces, thereby missing cherry picks performed without the \texttt{-x} option.
Importantly, patch ID-based detection is not conceptually tied to local analysis: it is a deterministic, per-commit computation that can technically be performed on the global commit graph, and its support over the entire Software Heritage archive is ongoing (cf.~\Cref{sec:threats}). It was, however, not yet available at the time of our experiments.
Within the scope of the evaluation of our approach, we therefore apply the same detection locally, on a clone of the repository of each fork potentially affected by a vulnerability, and propagate the resulting cherry-pick information exactly as the global-scale implementation will.
Since this correction is automatic and faithfully emulates the future global-scale behavior, evaluating our approach with it reflects the approach as deployed at scale; for transparency, we nevertheless also report the precision obtained without it.

Finally, an author evaluates all the remaining $\langle\mathit{fork}, \mathit{vulnerability}\rangle$ pairs to check whether an equivalent patch, not yet detected, had been applied in the fork, according to the following protocol:
 \begin{itemize}
     \item Read the CVE description.
     \item Read the diff of the commit that fixes the vulnerability.
     \item For each modified snippet (``diff hunk''), verify that it is indeed missing in the code base of the fork.
     \item If an alternative patch is found, mark the pair $\langle\mathit{fork}, \mathit{vulnerability}\rangle$ as false positive; mark it as true positive if not.
 \end{itemize}

\paragraph{Maintainer notification}
We contacted the maintainers of remaining forks identified as affected by one-day vulnerabilities.
This constitutes both responsible disclosure of potential security issues that we identified, and provides the opportunity to ultimately validate our findings by the most relevant experts that exist: the maintainers. Based on the maintainer responses, we ultimately classify presumed vulnerable forks as:
\begin{itemize} 
\item \emph{False positive}: the fork is not vulnerable for reasons we could not identify on our own, e.g., the vulnerability was not applicable to it or the patch was applied via means we did not detect;
\item \emph{True positive}: the fork is actually vulnerable.
\end{itemize}

\subsection{Results}

\subsubsection{Forks with unpatched head}
\begin{figure}[ht!]
 \captionsetup{}
     \centering
     \includegraphics[width=0.96\linewidth]{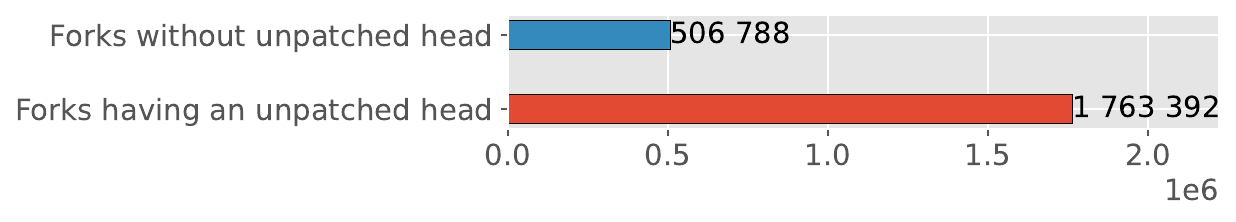}
     \caption{Forks having a presumed vulnerable commit in their history: proportion of forks having an unpatched head}
     \Description{Forks having a presumed vulnerable commit in their history: proportion of forks having an unpatched head}
     \label{fig:fig5}
 \end{figure}

We identify \unpatchedforkinheadshort presumed one-day vulnerable forks whose head commits remain unpatched. These account for 77.7\% of all forks previously identified as containing at least one vulnerable commit in their histories (\Cref{fig:fig5}), corresponding to \unpatchedpairinhead{} pairs of $\langle\mathit{fork}, \mathit{vulnerability}\rangle$.

\subsubsection{Filtering forks and vulnerabilities of interest}
\label{susubsec:downstream_filtering}

\begin{figure}[ht!]
    \centering
    \includegraphics[width=0.96\linewidth]{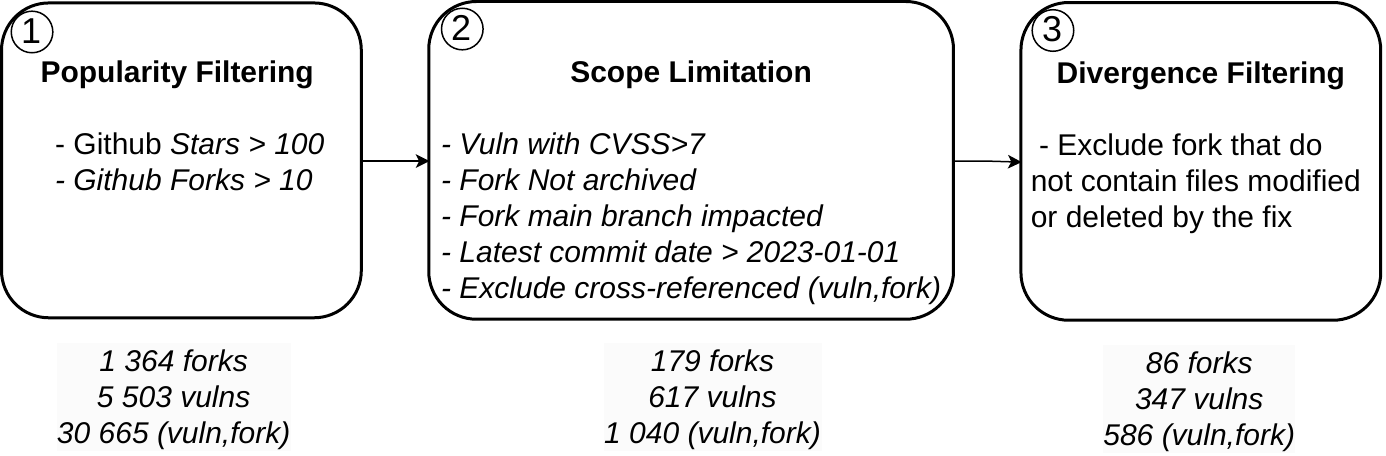}
    \caption{Result of the selection of forks and vulnerabilities of interest}
    \Description{Result of the selection of forks and vulnerabilities of interest}
    \label{fig:strict_filtering}
\end{figure}

\Cref{fig:strict_filtering} presents the detailed filtering workflow and reports, for each macro step, the number of forks and vulnerabilities retained after filtering.

The popularity filtering (cf. Stage 1 of \Cref{fig:strict_filtering}) yields 1364 forks out of 1.76 million, confirming that only a small fraction of forks with a presumed one-day vulnerability are popular. The second step, limiting the scope of the analysis (cf. Stage 2 of \Cref{fig:strict_filtering}) yields 1040 pairs  of $\langle\mathit{fork}, \mathit{vulnerability}\rangle$ out of \num{30 665}. Detailed numbers of each sub-filter of this stage are available in the reproduction package~\cite{DidYouForkgetItReproductionAnon}. The third step, filtering divergent forks (cf. Stage 3 of \Cref{fig:strict_filtering}), yields 586 pairs  of $\langle\mathit{fork}, \mathit{vulnerability}\rangle$ with 347 distinct vulnerabilities and 86 forks of interest. The divergence heuristic filters out 43\% of the pairs of $\langle\mathit{fork}, \mathit{vulnerability}\rangle$, meaning that the modified files associated with the vulnerability are not found in the fork. It confirms that, although sharing common history, forks can diverge significantly.

\subsubsection{Vetting of one-day vulnerable forks}
Now that we have selected forks and vulnerabilities of interest, we conducted the author and maintainer evaluation.

\begin{figure}[ht!]
    \centering
    \includegraphics[width=0.95\linewidth]{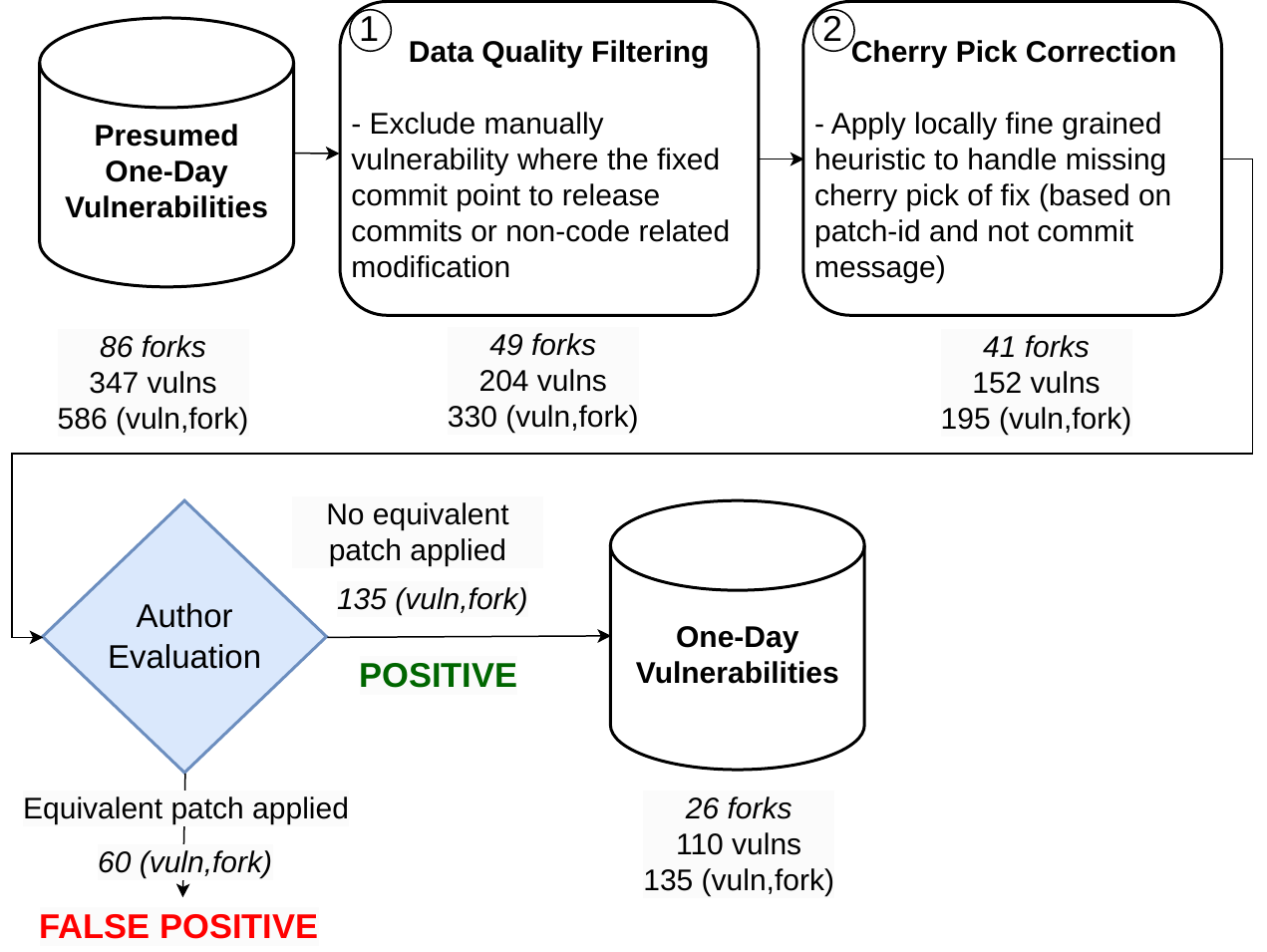}
    \caption{Author evaluation protocol}
        \Description{Author evaluation protocol}
    \label{fig:eval_author}
\end{figure}

\paragraph{Author Evaluation}
\Cref{fig:eval_author} presents the execution of the author evaluation protocol, reporting for each step the number of forks and vulnerabilities retained after filtering.

The data quality filtering (cf. \Cref{fig:eval_author}) yields 204 vulnerabilities out of the 347. 
We then applied the heuristic based on the patch ID of the fixing commits to identify and remove remaining cherry picks not yet handled at the global scale for implementation reasons (not a limitation of our approach). This heuristic yields 195 $\langle\mathit{fork}, \mathit{vulnerability}\rangle$ pairs associated with 152 unique vulnerabilities and 41 forks of interest. On average we observe \forkMean CVEs per fork with a median of \forkMedian, a first quartile of \forkqone, a third quartile of \forkqthree and a maximum of \forkMax. Two outlier forks are present and associated respectively with 100 and 8 CVEs. All of these pairs are then manually evaluated to determine whether an equivalent patch has been applied or not (cf.~\Cref{fig:eval_author}). This evaluation resulted in \AuthorEvalcountPOSITIVE positive and \AuthorEvalcountNEGATIVE false-positive  $\langle\mathit{fork},\mathit{vulnerability}\rangle$ pairs, corresponding to a precision of \AuthorEvalcountprecision.
This means that \AuthorEvalcountprecisionpercent\% of evaluated pairs considered positive are actually positive based on author evaluation. Given that the two outlier forks account for more than half of all evaluated pairs, we conducted a secondary analysis in which these outliers were excluded. This second analysis yielded a precision of \AuthorEvalWithoutOutliercountprecision, showing the limited impact of outliers.

For transparency, we also report the precision obtained without the patch ID-based cherry-pick correction, i.e., as achieved by the current prototype implementation alone: it decreases to 0.41.
This gap is entirely attributable to fixes cherry-picked without commit-message traces, which the prototype cannot yet recognize during global propagation. Since patch ID-based detection is technically applicable to the global commit graph and its integration into the Software Heritage graph is ongoing (cf.~\Cref{sec:threats}), we consider the corrected precision of \AuthorEvalcountprecision the relevant measure of the \emph{approach}, while 0.41 characterizes the current \emph{prototype} global propagation.

\subsubsection{Maintainer notification}
\begin{figure}[ht!]
 \captionsetup{skip=0pt}
    \centering
    \includegraphics[width=\linewidth]{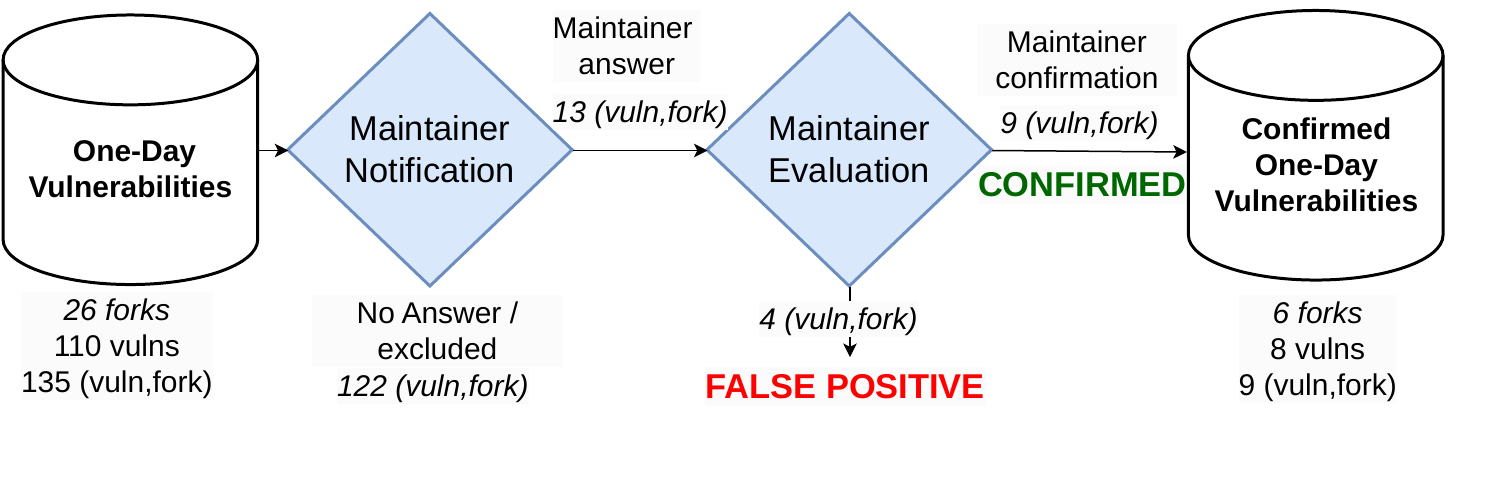}
    \caption{Maintainer evaluation protocol}
    \Description{Maintainer evaluation protocol}
    \label{fig:eval_maintainer}
\end{figure}

\Cref{fig:eval_maintainer} presents the execution of the maintainer evaluation protocol: for each step, the number of forks and vulnerabilities retained after filtering. Out of the \AuthorEvalcountPOSITIVE $\langle\mathit{fork}, \mathit{vulnerability}\rangle$ pairs, we excluded \OwnerNotificationcountEXCLUDED pairs because during the final due diligence before contacting maintainers, these forks were explicitly marked as deprecated or not intended for use (including some meant for educational purposes only). All the remaining fork maintainers have been notified. Among the candidates
for confirmation, here are a few notable examples: \texttt{sonyxperiadev/kernel} (the community Linux kernel fork by Sony, for Xperia Android devices), \texttt{panda-re/panda} (our initial example), \texttt{reMarkable/linux} (the Linux kernel fork for reMarkable eInk tablets), and \texttt{go-nv/goenv} (a fork of pyenv, adapted to the Go ecosystem).

We received \OwnerNotificationcountTOTALRESPONSE responses, with \OwnerNotificationcountAFFECTED confirmed one-day vulnerabilities and \OwnerNotificationcountNOTAFFECTED false positive $\langle\mathit{fork}, \mathit{vulnerability}\rangle$ pairs.
For the remaining ones, we are awaiting confirmation. The confirmed one-day vulnerabilities result in a maintainer-confirmation rate of \OwnerNotificationcountresponseprecision. This means that \OwnerNotificationcountresponseprecisionpercent\% of the
pairs for which we received a maintainer response were confirmed. We received two types of responses: (1) from active communities where the maintainers of the repositories quickly responded and fixed the vulnerability, (2) from less active forks, with communities lacking maintainer resources and struggling to keep up with upstream. An example of the first type of response is from \texttt{github.com/sonyxperiadev/kernel}, vulnerable to CVE-2021-45485 and CVE-2021-4154, whose maintainers fixed the vulnerability within days. The majority of the responses were of the second type, with maintainers acknowledging the vulnerable status of their fork, but lacking the resources to address it.

Our conclusion for this RQ is as follows:
\conclusion{%
\textbf{RQ:} \emph{Global history analysis} is effective in supporting the identification of downstream forks affected by one-day vulnerabilities, as exemplified by the \texttt{PANDA} and \texttt{Xperia} cases.
 Among the \AuthorEvalcountTOTAL evaluated pairs of $\langle\mathit{fork}, \mathit{vulnerability}\rangle$, \AuthorEvalcountPOSITIVE remain positive after author evaluation (precision of \AuthorEvalcountprecision).
Maintainers confirmed \OwnerNotificationcountAFFECTED one-day vulnerabilities out of \OwnerNotificationcountTOTALRESPONSE responses (confirmation rate \OwnerNotificationcountresponseprecision).

}

\section{Related work}
\label{sec:related}

\subsection{Detection of vulnerable dependencies}
The propagation of vulnerabilities through dependency graphs has garnered significant scholarly attention in recent years~\cite{Zhang_Does}, recognized as a critical threat to the security of the open-source software supply chain~\cite{enck2022top,10179304} and extensively investigated across major software ecosystems~\cite{zerouali2022impact, pashchenko_dep, li2021pdgraph}.
In parallel, tools like Renovate\footnote{\url{https://docs.renovatebot.com/}} and GitHub Dependabot~\cite{GitHubDependabot} have spurred a growing body of research on their potential and limitations~\cite{alfadel2021use, he2023automating, lin_vulnerabilities_2025}.
This paper shifts the focus from dependency-based reuse to fork-based reuse: while forks are popular in open-source, their role in the propagation of vulnerabilities remains comparatively unexplored.

\subsection{Detection of vulnerable fork versions}
Yi et al.~\cite{yi_blockscope_2023} identified the propagation of 116 vulnerabilities to 16 forks of two blockchain projects using code clone detection techniques, and found unpatched vulnerabilities in 13 out of 16 forks.
In comparison, our study operates at a much larger scale (thousands of projects, millions of forks) and is not restricted to forks of specific projects or application contexts; instead, we analyze the global commit graph of public code, using SWH as its proxy.

In more exploratory contexts, Reid et al.~\cite{Reid_21_tracing}, at the MSR 2021 Hackathon, used World of Code~\cite{ma2019world} to perform file history analysis: given a patch, they identify all previous versions of the modified files and consider them vulnerable, then identify forks whose HEAD contains such files.
This file-based approach cannot handle projects that have modified the impacted files.
In contrast, our approach is based on commit history analysis, allowing us to detect all commits that have the vulnerability-introducing commit in their ancestry without containing the corresponding fix.

\subsection{Detection of vulnerable software versions}
\label{sec:vuln_ver}

A large body of work detects vulnerabilities within specific software versions, which Lin et al.~\cite{lin_vulnerabilities_2025} classify into \textit{traditional} and \textit{neural-network-based} detection.
Traditional techniques include code clone detection at various granularities (text-, token-, tree-, graph-, and measure-based)~\cite{Kim_17_Vuddy,SajnaniSSRL16,ijcai/WeiL17,CuiHJFY21,LiZXJQH16}, binary code analysis for closed-source software~\cite{David_18_firmup,Chandramohan_16_Bingo}, and formal methods such as symbolic execution~\cite{10.1145/1455518.1455522}, fuzzing~\cite{chen2020savior}, and taint analysis~\cite{kang2022tracer}.
Neural-network-based approaches instead learn vulnerability patterns from curated datasets, using CNNs~\cite{russell2018automated}, GNNs~\cite{zhou2019devign}, or NLP-based models~\cite{kim2022vuldebert}.

\subsection{Hybrid history analysis}
\label{sec:hybrid-hist}
Closer to our approach, some techniques combine vulnerability detection with Version Control System history analysis to avoid analyzing every version individually.
Several build upon SZZ~\cite{sliwerski2005szz} and its variants, originally designed to identify bug-introducing commits from fixing commits.
For instance, V-SZZ~\cite{bao2022v} uses static analysis to identify bug-inducing commits and then traverses the Git history to identify vulnerable commits associated with version tags.
These algorithms address a fundamentally different problem from ours: they start from a commit known to fix a vulnerability and trace back the introducing commit, whereas we track forked repositories where, by definition of a one-day vulnerability, the fixing commit is not present yet. The challenges we address are therefore different: identifying all still-affected repositories (a scalability issue) and identifying impacted commits that may have different identifiers across forks (an identification issue), both of which we address by propagating OSV semantics across the global commit graph.
Wu et al.~\cite{wu2024vision} compare the execution time of six SZZ-derived approaches on 102 CVEs within 79 libraries and more than 12,073 library versions, showing that these take, on average, between 0.01 and 187 seconds per library version, and between 1.42 seconds and \num{17 771} seconds for analyzing all versions for a single vulnerability.
In our setting, we analyze \unpatchedpairinheadshort $\langle\mathit{fork}, \mathit{vulnerability}\rangle$ pairs across 1.7 million forks; applying prior approaches at this scale would require between 2.4 years and 30 thousand years under the same experimental conditions, and this estimate still assumes all forks are already identified and locally available---itself infeasible without a global model like ours.
We instead reuse an existing database (OSV) of introduction and fix commits and rely solely on commit history analysis, which enables our approach to operate at this scale; we acknowledge, however, that OSV inherits known quality issues~\cite{dong_vuln_report}, e.g., conservative declarations of the introduction commit, which SZZ-based approaches could help refine at a smaller scale.

Finally, several studies analyze patch propagation and CVE lifespans at the repository or branch level~\cite{li_large-scale_2017, liu_large-scale_2020}. Notably, Tan et al.~\cite{tan_understanding_2022} found that 80\% of CVE--branch pairs remain unpatched across 608 stable branches of 26 popular OSS projects, corroborating, at the branch level, the maintenance gap that our fork-level analysis also uncovers (\Cref{sec:discussion}).

\section{Discussion}
\label{sec:discussion}

\subsection{Ethical considerations}
\label{sec:ethics}
The vulnerabilities we study are one-day vulnerabilities: they are already public 
and patched in at least one repository of the fork ecosystem, but remain unpatched in the forks we identify. Our disclosure followed a 
responsible-disclosure process: for every fork we confirmed as affected, we privately notified its maintainers through the project's established contact channels, gave them the opportunity to remediate before any public release, and offered assistance in doing so (cf.~\Cref{sec:vetting}).
We contacted only project maintainers, and only about vulnerabilities in their own code; no end users were involved and no personal data was collected.

\subsection{Implication for development practices}

\paragraph{Forking is easy, maintaining is hard}
With the advent of collaborative platforms such as GitHub, soft-forking a code repository now takes a single click; maintaining a hard fork is not as straightforward, and in practice translates to only ``infrequent propagation'' of changes~\cite{Businge_22_maintenance_forks} from upstream to downstream repositories. Our analysis highlights a real maintenance issue: many vulnerabilities could be easily fixed, but are not, due to a lack of human resources in the downstream, forked project.
Limiting this risk requires a strategy for integrating upstream changes while limiting conflicts and manual intervention. Such strategies (periodic \emph{cherry-pick}, \emph{merge}, or \emph{merge rebase}), although used in practice\footnote{\url{https://web.archive.org/web/20250114052242/https://github.blog/developer-skills/github/friend-zone-strategies-friendly-fork-management/}}, are only covered in a few studies~\cite{Businge_22_maintenance_forks}, and ultimately require maintenance resources that many OSS projects simply do not have---underscoring the need for automated tools, not only to notify maintainers, like those we introduce here, but also to automatically apply fixes, a relevant direction for future work.

\paragraph{Limiting false positives}
Fork maintainers can limit false positives by adopting patch strategies our approach can recognize automatically, e.g., cherry-picks or merges, or by including the CVE ID in the patch commit when an alternative patch is used.
Beyond the divergence heuristic we propose in \Cref{sec:rq2_protocol} (based on the presence of files modified by a vulnerability patch in a fork), other heuristics could be designed, e.g., restricting the analysis to vulnerabilities impacting specific files: the maintainer of \texttt{sonyxperiadev/kernel} noted that, maintaining a fork of the Linux kernel for the ARM architecture, they are not impacted by x86-specific vulnerabilities even when that code is present in their fork.

\subsection{Threats to validity}
\label{sec:threats}
\paragraph{Conclusion validity} Our analysis is based on the global commit graph provided by Software Heritage, so our approach only applies to forks whose commit history is publicly accessible, excluding closed-source forks---though it could similarly be used on the closed-source commit graph of an organization's codebase.

\paragraph{Reliability} Our results can be reproduced with the same input data. The manual analysis protocol is susceptible to human error, though we mitigated this risk with a rigorous protocol culminating in validation by project maintainers.

\paragraph{Internal validity}
Our approach (cf. \Cref{fig:strict_filtering}) relies on GitHub popularity metrics to identify forks with relevant user bases potentially affected by high-severity vulnerabilities. Different filtering conditions could be designed based on other platform-specific popularity metrics or intrinsic metadata~\cite{maj_stars}.
The divergence filter is deliberately conservative: it excludes a $\langle\mathit{fork}, \mathit{vulnerability}\rangle$ pair as soon as a single file modified by the fix is absent from the fork. This can drop genuinely vulnerable pairs (false exclusions), and, conversely, the presence of the modified files does not guarantee that the vulnerable code path is still reachable. This heuristic therefore trades recall for precision, consistent with our goal of surfacing high-confidence one-day vulnerabilities for disclosure.

\paragraph{Construct validity}
As detailed in \Cref{sec:vetting}, the current prototype implementation of global propagation detects cherry-picked fixes only through commit-message traces (\texttt{"cherry picked from commit~\ldots"}, cf \texttt{git cherry-pick}\footnote{\url{https://git-scm.com/docs/git-cherry-pick}}). Consequently, the global counts reported in \Cref{fig:fig5} have an uncorrected precision of 0.41, as measured on our evaluation sample. The reported corrected evaluation additionally uses patch IDs to detect cherry-picked fixes without message traces. Although applied locally to the evaluation sample, this deterministic technique can be implemented globally and therefore estimates the precision of the complete approach. The confirmed vulnerabilities are unaffected because this correction preceded the author evaluation, and archive-wide patch-ID computation is currently being integrated into Software Heritage. In addition, following OSV semantics, a merge may be labeled patched even if conflict resolution discards the fix. 

\paragraph{External validity}
Our analysis depends on the Software Heritage archive and can be impacted by inconsistencies in the archived data, as well as by the quality of the vulnerability database~\cite{dong_vuln_report}.
Incorrect ``fixed'' commits, which do not contain the actual fix, such as those filtered in \Cref{sec:vetting}, can prevent the evaluation of the fork impact.
However, this does not necessarily mean that the propagation of such events is incorrect, as they can effectively be treated as ``last\_affected'' events.

\section{Conclusion}
\label{sec:conclusion}
In this paper, we leverage the global commit graph to track, at a large scale, one-day vulnerabilities (i.e., known but unfixed vulnerabilities) across forks of open-source projects.
More importantly, our analysis highlights the opportunity to leverage the shared development history between upstream projects and their forks to develop automated tooling that can reinforce the security of the global OSS supply chain.

Experimentally, by tracking one-day vulnerabilities associated with \upstreamrepo initial repositories, we identified  \impactedcommitinforkshort presumed-vulnerable new commits (missing from the original upstream repository) in forks and \unpatchedforkinheadshort potentially unpatched forks.
We validated our approach by manually vetting unpatched forks, ultimately obtaining confirmation by upstream maintainers of \OwnerNotificationcountAFFECTED high-severity one-day vulnerabilities in actively used forks.

This work paves the way to semi-automated analyses benefiting fork maintainers, users, and security auditors, by notifying them of potential one-day vulnerabilities in relevant code repositories. It also demonstrates the need to study vulnerabilities induced by code reuse practices beyond explicit dependencies, a pressing concern under the Cyber Resilience Act~\cite{cra2024}, which requires that products be placed on the market free of known exploitable vulnerabilities, turning systematic one-day vulnerability analysis from good practice into a compliance obligation.

\begin{acks}
  This work was made possible by Software Heritage, the universal source code archive: \url{https://www.softwareheritage.org}.
  This work was supported by France Agence Nationale de la Recherche (ANR), program France 2030, reference ANR-22-PTCC-0001.
\end{acks}

\bibliographystyle{ACM-Reference-Format}
\bibliography{main_reducted}
\end{document}